# Security Model For Service-Oriented Architecture

Oldooz Karimi, MSc

Department of Computer Engineering, sofiyan Azad University, sofiyan, Iran,

`oldooz_karimi@yahoo.com`

**Abstract**.

*In this article, we examine how security applies to Service Oriented Architecture (SOA). Before we discuss security for SOA, lets take a step back and examine what SOA is. SOA is an architectural approach which involves applications being exposed as "services". Originally, services in SOA were associated with a stack of technologies which included SOAP, WSDL, and UDDI. This article addresses the defects of traditional enterprise application integration by combining service oriented-architecture and web service technology. Application integration is then simplified to development and integration of services to tackle connectivity of isomerous enterprise application integration, security, loose coupling between systems and process refactoring and optimization.*

## 1 Introduction

It is tempting to launch into a description of SOA Security without first asking "Why?" Why apply security to SOA? One obvious answer is to protect the SOA infrastructure against attack. This is a valid reason, but there are also enabling, positive reasons for applying security to SOA, such as the ability to monitor usage of services in a SOA. We begin by examining the attacks against SOA technologies, both EAI and ESA. Abstract: Interoperable software architecture requires interoperable security mechanisms. Security is frequently looked at as a black art, but in reality the core concepts of security - knowing your assets and designing for failure - are just good engineering practices. This article focuses on applying those practices to service-oriented solution design with an emphasis on considerations raised by authentication, authorization, auditing, and assurance.

___________

## 2  SECURITY CHALLANGES

Over time, application architecture has become more complex progressing through mainframe centric, client server, distributed computing, loosely coupled architecture, to Service Oriented Architecture (SOA). With each change in application architecture security has become more complex.

Consider the needs for a loosely coupled architecture such as Enterprise Application Integration (EAI), having the goal of building composite applications from standalone applications with the use of Message Oriented Middleware (MOM), an integration broker and application adapters. In essence, EAI bypasses user-based security (e.g. a GUI sign-on) and creates new system-to-system based security. Given the assumption that the standalone applications are secure, integrating these applications via APIs or direct database access will present the following new security requirements:

* Authentication to the MOM server
* Link level encryption or network segregation of MOM messages
* Security of credentials passed to the standalone systems from the EAI broker (e.g. simulating a user sign-on, database or API credentials)
* Security of the EAI infrastructure (against unauthorized access, denial of service attacks, tampering with information flows)

EAI introduces a level of complexity to security that is often overlooked. Practices such as hard coding security credentials into integration broker logic and transmitting critical information in clear text using MOM over the network are common occurrences. Consider now the next architecture shift to SOA where consumers can dynamically locate service producers. The inherent benefit if SOA is the loose coupling between the producer and the consumer, which eases the construction of component based solutions and promotes abstraction. To preserve loose coupling, security must also be implemented as a service - to avoid tightly bound security and thereby tight binding of the services themselves. SOA has gained commercial momentum since the introduction of the standards based approach of Web Services. However, the use of Web Services presents yet another set of security challenges as follows:

* The typical transport protocol is HTTP(S) which is open on most firewalls
* Both users and services are distributed - often times over the Internet
* Abstracting legacy systems as Web Services can introduce multiple underlying messaging protocols and security mechanisms - Web Services in this case are often exposed via the ESB with similar security risks as EAI
* Data exchange via XML within a SOAP envelope introduces a new set of vulnerabilities - such as XML denial of service exploits of XML parser vulnerabilities and buffer overflows

The complex security requirements of SOA and specifically Web Services are driving a new set of security standards as follows:

* SAML Security Assertion Markup Language SAML is an XML framework for exchanging authentication and authorization information.

* WS-Federation Web Services Federation Language This specification defines mechanisms to allow different security realms to federate by allowing and brokering trust of identities, attributes, authentication between participating Web services.

* WS-Security Describes enhancements to SOAP messaging to provide quality of protection through message integrity, message confidentiality, and single message authentication.

* WS-Secure Conversation Defines extensions that build on WS-Security to provide secure communication. Specifically, it defines mechanisms for establishing and sharing security contexts, and deriving session keys from security contexts.

* WS-Security Policy An addendum to WS-Security. Indicates the policy assertions for WS-Policy which apply to WS-Security.

* WS-Trust Defines extensions that build on WS-Security to request and issue security tokens and to manage trust relationships.

* XML-Encryption Specifies a process for encrypting data and representing the result in XML.

* XML-Signature Specifies XML digital signature processing rules and syntax.

These security standards are at varying degrees of completion and commercialization and in some cases competing standards have been proposed. Where the standards are incomplete security software vendors have implemented proprietary solutions. There are also proprietary solutions for such security problems as:

* Workflows to provision and terminate users

* Single Sign-on

* Synchronization of disparate directories

Given the complex requirements for SOA security, the immaturity of Web Services security standards and the requirements for legacy system security, there is a strong technology case to consider commercial security packages as the infrastructure for identity management within the SOA.

## 3. SECURITY APPLIANCE ARCHITECTURE

SOA security appliances implements security as a service through a hardware-based gateway and XML proxy that can parse, filter, validate schema, decrypt, verify signatures, access-control, transform, sign and encrypt XML message flows. The security appliances is a server side security gateway that allows for all keys and tokens used to provide integrity and confidentiality to services exposed through the gateway to be managed at one point, the appliance. That means that clients invoking any number of services exposed through this appliance need only be configured to trust keys or tokens from this single gateway, rather than keys and tokens from each service. The appliance needs to be configured to trust keys and tokens from each client. Security appliances should implement the following features:

* Comprehensive security - All XML and web services security functions in one device
* Web services access control via new technologies (like SAML, XACML, WS-Security) and existing systems (like LDAP and SSO) to control access to applications
* Centralized configuration and policy management
* Performance - purpose-built to secure without degradation
* XML-based agility - future-proof for changing standards, policies, partners
* Appliance-based - drop-in device secures multiple applications at once
* Easy integration - interoperates with and augments existing security systems
* XML transformation - includes XPath/XSLT acceleration features

## 4. SERVICE ORIENTED ARCHITECTURE

Business process and solutions should be designed and developed using a Service Oriented Architecture (SOA) approach. The SOA approach is considered best practice and is used by many organizations to improve their effectiveness and efficiency in the provision of IT services. SOA is defined by OASIS (www.oasis-open.org) as:

'A paradigm for organizing and utilizing distributed capabilities that may be under the control of different ownership domains. It provides a uniform means to offer, discover, interact with and use capabilities to produce desired effects consistent with measurable preconditions and expectations.' OASIS (Organization for the Advancement of Structured Information Standards) is a not-for-profit, international consortium that drives the development, convergence and adoption of e-business standards. SOA brings value and agility to an organization by encouraging the development of 'self-contained' services that are re-usable. This, in turn, promotes a flexible and modular approach to the development of 'shared services' that can be used in many different areas of the business. More and more organizations are converting business processes to common 'packaged services' that can be used and shared by many areas of the business.

Wherever possible, IT service provider organizations should use the SOA and principles to develop flexible, re-usable IT services that are common and can be shared and exploited across many different areas of the business. When this approach is used, it is essential that IT:

· Defines and determines what a service is

· Understands and clearly identifies interfaces and dependencies between services

· Utilizes standards for the development and definition of services

· Uses common technology and tool-sets

· Investigates and understands the impact of changes to 'shared services'

· Ensures that SOA-related training has been planned and achieved for the IT people in order to establish a common language and improve the implementation and support of the new or changed services.

When SOA principles are used by the IT service provider organization, it is critical that an accurate Service Catalogue is maintained as part of an overall Service Portfolio and Configuration Management System (CMS). Adopting this approach can significantly reduce the time taken to deliver new solutions to the business and to move towards a Business Service Management (BSM) capability. The Service Catalogue will also show the relationship between services and applications. A single application could be part of more than one service, and a single service could utilize more than one application.

## 5. BUSINESS SERVICE AND IT SERVICE

A business process can be distributed across technologies and applications, span geographies, have many users, and yet still reside in one place: the data centre. To integrate business process, IT frequently employs bottom-up integration, stitching together a patchwork of technology and application components that were never designed to interact at the business process layer.

What began as an elegant top-down business design frequently deteriorates into a disjointed and inflexible IT solution, disconnected from the goals of the business.

An improved strategy for engaging at the business process layer is focusing on modelled abstractions of business activities. These focal points, called business services, represent business activities with varying degrees of granularity and functionality. A business process, for example, may be represented as a single business service or a collection of business services . A business service can represent a composite application or a discrete application function.

It may represent a discrete transaction or a collection of supporting fulfillment elements. In all cases, it exists in the domain of the business.

A business service is defined by the business. If IT provides a service to the business, but the business does not think of the service in any business context or semantics, then it is an IT Service. By considering services as a system for creating and capturing value, regardless of sourcing or underpinnings, the line between IT Services and business services begins to blur. Instead, each can be thought of as different perspectives across a spectrum. Again, the decision to adopt a business or IT perspective depends on the context of the customer.

When this notion is combined with other seemingly unrelated service-oriented technologies and concepts, their relationships can be illustrated in the chart shown in Figure 1.

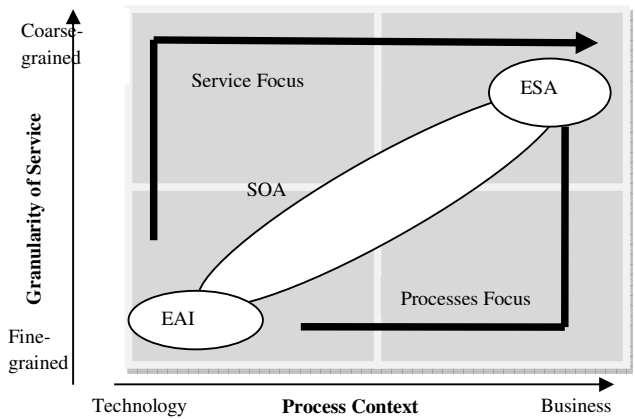

Figure 1 : Service perspective

Figure 1 states that all services, whether they are IT Services, business services or services based on Service-oriented Architecture (SOA), Enterprise Services Architecture (ESA) or Enterprise Application Integration (EAI), are members of the same family. They may differ by granularity (fine versus coarse) or by context (technology versus business). They each provide a basis for value and require governance, delivery and support. ITSM and BSM are each perspectives on the same concept: service management.

## 6. VIRTUE AND LIMITATIONS OF EAI

In the original sense the term EAI described only the concept of semantic integration, but since the evolvement of systems covering this domain the term has also been used to name the systems itself.

To subsequently show how semantically enabled SOA are applied in EAI scenarios we first depict the functionality constituting these EAI systems in the following paragraph.

### A. EAI Concepts and Functionality

We can differ between internal and external integration. Internal integration, often referred as intra-EAI, specifies the automated and event-driven exchange of information between various systems within a company. Another commonly used term for it is 'Application to Application'-Integration (A2A) [7]. External integration, referred as inter-EAI, specifies the automated and event-driven information exchange of various systems between companies. In recent publications the second integration type is commonly referred to as B2B integration [7].

EAI systems provide different types of integration levels explaining the various dimensions of the integration tasks. Literature identifies the following three layers common to EAI systems [14, 9].

### Process Layer

Within the process layer we differ between components for process modelling and process enactment support. The purpose of process modeling is to produce an abstraction of a process model called workflow type [10] that can be either used for improved human understanding of operations within a domain or to serve as the basis for automated process enactment. In the case of EAI it is used for the latter. Process enactment refers to the proactive control of the entire process from instancing a predefined workflow type all the way to its completion. When talking about process models in EAI a clear separation between private and public processes has to be made. This separation is well established in research [6] and different business process modelling standards (i.e. BPEL [1], WSCI [2]) incorporate it. It is the key to provide the necessary isolation and abstraction between the organisation's internal processes and processes across organisations

### Transformation Layer

This layer ensures the proper routing and transformation of messages between the applications to be integrated. Transformation refers to a process of selecting, targeting, converting and mapping data so that it can be used by multiple systems [14]. The transformation addresses the mismatch of data either at the lower-level of data type representation or at the higher-level of mismatched data structures. Mismatched data types may arise when two services use different binary representation for some data type. Dissimilar data structures on the other hand involve two different structures to represent the same body of data.

### Transportation Layer

The Transportation Layer provides the system- and platformindependent communication between the integration tool and the participating applications. It consists of a common protocol layer and adapters which transform external events in messages and vice versa [14].

Most EAI systems support some sort of asynchronous transport layer, either proprietary or open ones. For instance, many leverage IBM WebSphere MQ or more open systems such as Java Messaging Service (JMS) or the Object Messaging System (OMS) [33].

**B. Service-Oriented EAI Architectures**

SOA is a set of independently running services communicating with each other in a loosely coupled manner via event-driven messages. Although the concepts behind SOA were established before Web Services came along and a service within a SOA is completely independent of the concept of aWeb Service, current SOA architectures employ them [11];Web Services naturally implement the philosophy of a SOA by using lightweight protocols based on widely accepted standards.

Established technologies for SOA include for example the Common Object Request Broker Architecture (CORBA)4 administered by the Object Management Group (OMG). It is a vendor-independent architecture that applications based on different programming languages can use to work together over networks. In contrast to Web Services which are mostly based on SOAP/HTTP, services in CORBA typically communicate via the IIOP (Internet Inter-ORB Protocol). The second major differences is the tight coupling of two CORBA services in comparison to the loose coupling between Web Services. In CORBA objects are shared between components, whereas Web Services communicate primarily over message.

The difference in the SOA approach to traditional EAI lies primarily in the way applications are seen in the architecture. The applications functionality is broken down into modules with well defined standardised interfaces. SOA treats services as a means of providing one functionality in a whole business process. Applications have to provide the standardised interface themselves in order to be integrated. If they lack such an interface, it is still required to wrap their functionality to a well-defined Web Service interfaces in order to participate in interactions with other applications.

This wrapping process creates an abstraction layer that hides all the complex details of the application similar to that achieved with traditional EAI adapters. The difference lies in the standardization of the interface.

At a conceptual level, SOA is composed of three core pieces [13]:

Service registry: It acts as an intermediary between providers and requesters. Most of these directory services categorise services in taxonomies.

Service provider: The Service Provider defines a service description and publishes it to the service registry.

Service requester: The service requester can use the directory services' search capabilities to find service descriptions and their respective providers.

The three activities the service requester and provider in a SOA as depicted in figure 2 can perform are [13]:

Publish: The service provider has to publish the service description in order to allow the requester to find it. Where it is published depends on the architecture.

Discover: In the discovery the service requester retrieves a service description directly or queries the service registry for the type of service required.

Invoke: In this step the service requester invokes or initiates an interaction with the service at runtime using the binding details in the service description to locate, contact and invoke the service.

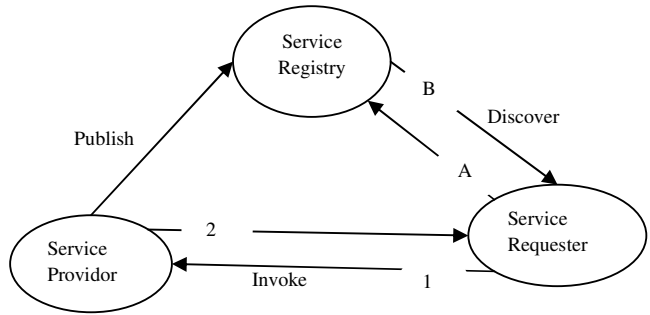

Figure 2: Activities in a SOA cp. [12]

## C. SOA in EAI and its weakness

In the following we show how traditional SOA addresses the three functional layers of EAI and identify its limitations which are mainly caused by the absence of semantics in the service descriptions.

### Process Layer

The technology of choice within SOA EAI solutions to implement private business processes are Workflow Management Systems (WfMSs) [10]. As mentioned above WfMSs used for EAI have to allow both the definition of workflow types and the execution of workflow instances. Traditional SOA implementations in the EAI domain apply to a growing extent XML based Business Process Modelling Standards like BPEL [1], WSCI [2], BPSS [10] etc. Microsoft BizTalk is a prominent SOA incorporating one of these standards, namely BPEL. BPEL clearly separates private and public process models. It introduces the concept of executable process for private processes and abstract process for public processes.

If it is possible at all to model the public processes, this definition is not executable, regardless of what standard is chosen in the process layer. One has to incorporate the public process in the execution of its private process [6]. The behaviour of the public process of two services exposed by different companies have to match to establish a communication. This leads to the necessity to define one process model for every partner the organisation is conducting business with. An alternative approach is to model the executable private process separately from the executable public process and define binding to relate both [6]. However, WfMS used within SOA does not have the concepts of dynamic binding of public and private processes. As Bussler [6] describes, WfMS assume that workflow instances execute in isolation. Two top level workflow instances (in this case the private and the public process) can not be related by current WfMSs since no modelling concepts are available for this functionality.

### Transformation Layer

Since SOA based on Web Services use XML as their way of defining the data structure of messages, transformation does not have to deal with mismatches between e.g. hierarchical structures or fixed-length fields. However in XML the same data item may be modelled as an attribute of an existing element in one document and as a subelement in another XML representation. In current SOA document transformation becomes a functionally exposed mapping subprocess. Since current standards like XML and XML Schema only solve the mismatch on the syntactical and structural level; solving the mismatch on the semantic level is usually handled on a case-by-case basis. If, for example, two services use RosettaNet as their data structure, both trading partners have to transform RosettaNet to whatever internal data structures they use. Transformation maps are used to process and convert the content and structure of any source information based on its XML schema representation into any target document format [15].

This solution requires a mapping between every two different XML schemas before an interaction between the respective Web Services can be set up. A widely used standard protocol for creating and saving this map pings is XSLT.

**Transportation Layer**

SOAP as the protocol used for exchanging structured information in a Web Service enabled SOA can be used with a variety of transport protocols such as HTTP, SMTP, and FTP.

SOA applying Web Services assumes a WSDL compliant interface of the participating applications. If the application does not natively support a WSDL interface, adapters are used to transform external events into messages and vice versa.

Both the transport protocol and the adapter integration is not further discussed.

## 7. SoAS FOR SECURITY

Service oriented architectures (SOA) is the natural evolution of client-server computing. Services are readily created by anyone and provisioned to users by making them available at an IP address on a port. This is core to the design of the Internet, where anyone can throw up a service and provide it to anyone who wants to use it. But everyone doing what they want does not cut it in most enterprise environments today. They generally want a little bit more control over how people spend their time and what they spend it on. So SOA in enterprises today is largely about (1) herding the cats before the enterprise gets out of control and (2) coming to group consensus about the shared services that will be most helpful, the protocols for using them reliably, and how they will continue to operate and be cared for over time.

In the security arena, there are emerging SOA capabilities for things like authentication services and audit trail collection. In the case of an SOA for security, unlike a new benefits application or an internal Web site, the security element (SE) can rapidly become a critical component of enterprise operations. Like a domain name server (DNS) that translates the name of a site into the numerical Internet Protocol (IP) address value necessary to send packets to that service, an SE that is useful becomes a critical component in the operation of large numbers of enterprise systems. And like the risk aggregation effects of DNS servers whose failure results in rapid collapse of operations, the failure of an SE can have serious negative consequences to the enterprise that uses it.

## 8. SECURITY SERVICES

Each progression in distributed computing - from object-orientation to component-based design and now to service-orientation - has introduced unique security considerations. Objects and components use similar binary runtimes, but when building services as Web services, we can no longer rely on binary controls for security (Table 1).

Table 1: Different distributed programming paradigms introduce different security considerations.

|  | Object-orientation | Component-Based Design | Service-Orientation With Web Services |
|---|---|---|---|
| Paradigm | Abstract models(objects) used to bundle data and methods | Technology-specific implementation model for distributed programming | Services designed as autonomous and standardized programs with an emphasis on reuse |

| Security Implications | Vender or implementation provides security mechanisms,authentication. Authorization,audit for object processing runtime(example:Java authorization services in JDK) | Component model provides implementation specific security models(example:cont ainer security in EJB) | Interoperable industry security standards deal with message security, identity federation, and service security(example: WS-Security) |
|---|---|---|---|

The primary security functions required by most systems are:

• authentication

• authorization

• auditing

• assurance

In SOA, these four functions can correspond to processing functionality provided by reusable utility services, which I'll refer to as "security services." Figure 3 shows how security services are needed to mediate communication between a subject and its objects - or, in the SOA world, between the service provider and its requesters (or consumers).

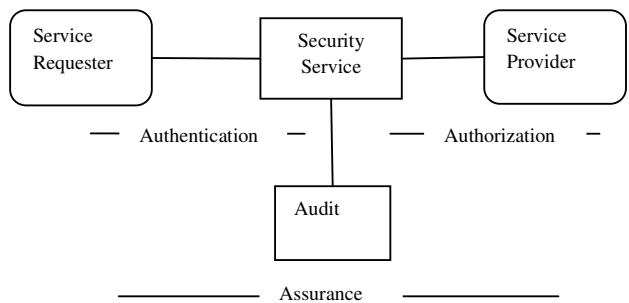

Figure 3: How security services can be positioned as intermediaries between service requester and provider.

A base SOA use case consists of service requesters, service providers, and message exchange patterns. It is within the message exchange where the authentication, authorization, audit, and assurance services add true value.

From a development perspective, the authentication and authorization services are frequently combined at runtime, but when classifying them as different service models, a logically decoupled service layer view is established.

Let's take a closer look at each of the security-centric service models:

• Authentication is concerned with validating the authenticity of the request and binding the results to a principle. This is frequently a system-level service because it deals with the processing of system policies (such as password policies) and implementing complex protocols (like Kerberos). This warrants a separate service because authentication logic is generally not valuable (or reusable) when intertwined with other application logic.

• Authorization, on some level, is always enforced locally, close to the thing being protected. In SOA, this thing is the service provider. While coarse-grained authorization can be implemented at a global level, finer grained authorization requires mapping to the service and its operations. From a design perspective, authorization should be viewed at both system and service levels (the latter always being enforced locally).

• Audit services provide detection and response features that serve to answers questions around what digital subjects performed what actions to what objects.

• Assurance services essentially exists as a set of system processes that increase the assessor's confidence in a given system.

## 9. Conclusion

Incorporating security requirements during early stages of software development will improve the important aspect "Security" of SOA based Information Systems. We have facilitated the business process expert in modelling the security requirements along with the business process model.

There is no perfect security solution, there is only the management of security risk that relies on judgment and prioritization, driven by assets and values. Security is contextual and has a form factor that must adhere to that which the supporting mechanisms can protect. Risk is increasingly engendered in data and effective security mechanisms adhere to data to provide the necessary level of protection. When SOA security standards are properly leveraged, the potential is there to create entirely new and robust service-oriented security architectures.

**oldooz karimi** holds a BSc on Software Computer Systems and an MSc in the Software Computer sign, both from the University of AZAD. l. Her research interests include Service , SOA, Enterprise, Enterprise architectures and Security. She has written extensively in these areas. She can be contacted on oldooz_karimi@yahoo.com.

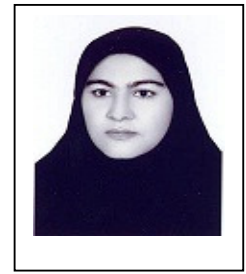